\documentclass[%
 reprint,
 amsmath,amssymb,
 aps,
 pre,
 floatfix,
]{revtex4-1}

\usepackage{graphicx}
\graphicspath{{images/}}
\usepackage{dcolumn}
\usepackage{bm}

\usepackage{enumitem}

\begin{document}

\title{Spectrum of chain oscillation in Poiseuille flow}
\author{M. Guskova$^{1,2}$}
\author{L. Shchur$^{1,2}$}
 \email{lev@landau.ac.ru}
\affiliation{%
 $^1$ Landau Institute for Theoretical Physics, 142432 Chernogolovka, Russia \\
 $^2$ HSE University,  101000 Moscow, Russia
 }%

\date{\today}

\begin{abstract}
We simulate solid particles moving in the two-dimensional channel with the Poiseuille flow. We found that the collective chain excitation emerges with the increasing number of particles in the chain. We measured the spectrum of the chain oscillations varying the width of the channel and found the spectrum intensity became sharper for the more significant confinement ratio.  The simulations were done using the Immersed boundary and Lattice Boltzmann methods.  We compare our results with the experiments of the drop movement in the quasi-two-dimensional channel and with the simulations of other groups. The paper's main result is that the combination of the velocity gradient in Poiseuille flow and the proximity of the particle to the wall can induce the collective excitations in the chain of the particles moving in the two-dimensional channel. 
\end{abstract}
\maketitle

\section{Introduction} 
\label{sec-intro}

Collective motion of the particles at the low Reynolds number in the channels filled with the fluid is subject of the intensive research connected in particular with the microfluidic devices~\cite{Doinikov-2020,Narayan-2020,Coppola-2021}. It was very intriguing that experiments with the drop chain movement in the quasi-two-dimensional channel demonstrate the nontrivial spectrum of the collective oscillations of the drop chain connected with the quasi-two-dimensional nature of flow~\cite{BTBZ-2006,BBZT-2007}. The water drops are squeezed between two parallel plates, and their velocity is slower than the surrounding oil due to the friction with the plate.  The spectrum was attributed to the dipolar potential due to the difference in the fluid and chain velocities, and it is possible to replace drops with the dipoles effectively. Based on that brilliant idea, the spectrum is calculated analytically and compared well with the experimentally measured spectrum~\cite{BBZT-2012}.

In the paper, we examine the nature of the spectrum oscillations of the solid drops (disks) moving in the two-dimensional Poiseuille flow. We vary the number of disks of radius $R$ in the channel with width $W$, and vary the value of the confinement ratio $\gamma=2R/W$. Several well-known analytical results predict the fate of the lonely particle in the Poiseuille flow.
It was argued in Ref.~\cite{Rubinow-1961} that the single disk is moving transversally to the streamlines due to the Magnus-like effect, and the lift force was calculated. In the Ref.~\cite{Saffman-1965} the lifting force on a sphere was calculated in the unbounded shear flow at very low Reynolds numbers. The transversal migration of the particles in the channel was found experimentally~\cite{Segre-1961,Goldsmith-1961,Karnis-1963}. The migration details do depend on the many factors, and particles may migrate to the channel axis, migrate to the wall, or stay at some finite distance from the axis and wall. No doubt both theoretically and experimentally that shear flow may lead to the transverse to the streamlines migration of the moving particles~\cite{Diamant-2009}.

In the paper, we investigate the emergence of the collective oscillations of the particle chain in the Poiseuille two-dimensional flow. We simulate the particle chain movement varying the channel width and the number of particles in the chain. The lonely particle in the Poiseuille velocity profile goes to the steady-state position asymptotically. It is due to the force, which acts on the drop perpendicular to the channel axis, and oscillations decaying due to the viscosity of the fluid~\cite{Saffman-1965}. We found that the particle chain spectrum appears with a number of drops larger than approximately ten. The spectrum nature is indeed due to the collective motion of the drops in the chain, which authors of Letter~\cite{BTBZ-2006} named the one-dimensional flowing crystal. In our computing experiment, we attribute the spectrum to the hydrodynamic interaction between particles -- the waves are generated in a fluid by particle movement and rotation and reflected from the channel boundaries, thus influents other particles in the chain.

For the simulation of the flowing neutrally buoyant disks in the two-dimensional channel filled with the fluid, we use immersed boundary method~\cite{Inamuro-2012} with the Lattice Boltzmann method~\cite{Krueger-book}.

\section{Setup and Model}
\label{sec-model}

In this section, we describe the model we use in simulations.  We do not explicitly consider the transient processes associated with initializing the pumping in the laboratory experiment and simulate the steady regime only.

\subsection{Geometry of simulation}
\label{sec-geometry}

The channel with the length $L$ and width $W$ is filled by the fluid
 from the left side. We start with the Poiseuille initial profile of the velocity and keep the left and the right boundary conditions with the Poiseuille velocity profile 

\begin{equation}
u_{x=0}=u_{x_L}=u_{max}\left( 1-\left( \frac{2y}{W} \right)^2 \right), 
\label{eq:profile}
\end{equation}
where $y$ is the distance from the axis, $y\in(-W/2,W/2)$. The hard disks of radius $R$ are placed equidistantly along the axis with the spacing between centers $a$. The initial velocity $v_i$ of the disks $i = 1, 2, \ldots, N$ is equal to the velocity of the flow at the axis $v_i(t=0)=u_{max}$.

\begin{figure}
	\begin{center}
	\includegraphics[width=1\columnwidth]{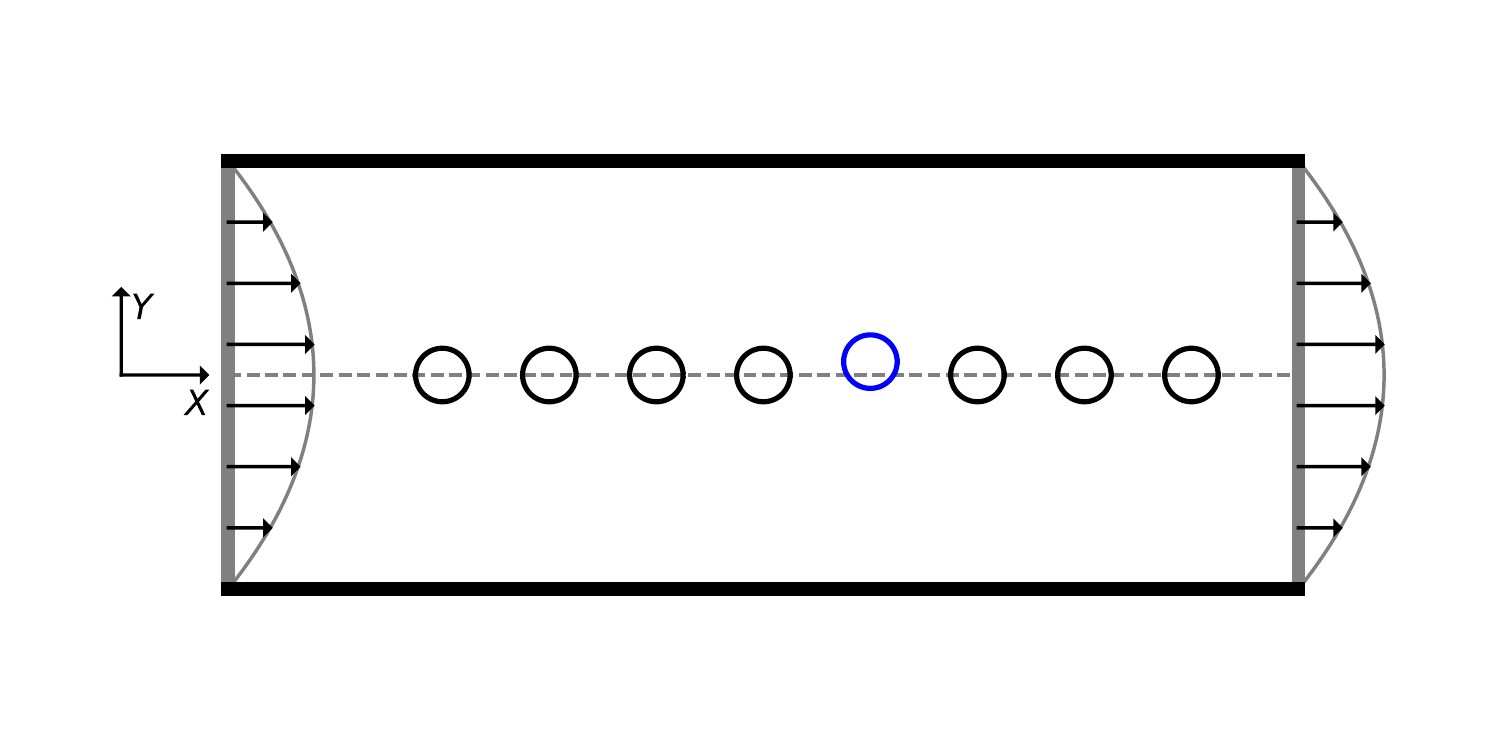}
	\caption{The initial position of the disks with $N = 8$. The fifth disk is shifted up by $R/4$. }
	\label{fig:initial}
	\end{center}
\end{figure}

The initial symmetrical displacement of the particles is stable during the simulations. Disk oscillations can emerge only due to the broken displacement symmetry. We consider the following initial conditions realizing possible disturbances of the symmetry, which may trigger chain oscillations: 1) shifting one on the disks from the axis, and 2) random moderate shifting of all disks. More details are present in Section III. The example of the initial condition is shown in Figure~\ref{fig:initial}, with the middle disk shifted in the up direction of the $y$-axis.

\subsection{Computational model}

The two-dimensional flow of fluid through the channel simulated using lattice Boltzmann method (LBM)~\cite{Krueger-book}. The distribution function $f_i (\vec{x}, t)$ represents the density of particles with velocity $\vec{c_i} (i=0,1,\ldots, 8)$ at the grid position $\vec x$ at the time $t$. We use D2Q9 with nine velocities $\vec{c_i}$ pointing to the four angles of the square around the grid position $\vec x$, to the four sides of the square, and to the center (zero velocity). The channel width is $W\Delta x$ with $\Delta x$ grid spacing, and the channel length is $L\Delta x$; $W$ and $L$ are integers. We use the bounce-back boundary conditions on the channel sides and establish the Poiseuille velocity profile at the left and right boundaries (see Eq. (\ref{eq:profile})). 

Distribution functions changes using the conventional LBM equations with the relaxation term proportional to the difference between the current $f_i (\vec{x}, t)$ and equilibrium  distributions $f_i^{eq} (\vec{x}, t)$~\cite{Succi-book}

\begin{equation}
f_i(\vec{x}+\vec{c_i}\Delta x,t+\Delta t)=f_i(\vec{x},t)-\frac{\Delta t}{\tau}\left[ f_i(\vec{x},t)-f^{eq}_i(\vec{x},t)\right]
\end{equation}
with the discrete equilibrium distribution function
\begin{equation}
	f_i^{eq}(\vec{x}, t)=w_i\rho\left(1+\frac{{\vec u}\cdot {\vec c}_i}{c_s^2} +\frac{({\vec u}\cdot {\vec c}_i)^2}{ c_s^2}-\frac{{\vec u} \cdot {\vec u}}{ 2c_s^2}  \right), \label{feq}
\end{equation}
where weights $w_i$ correspond to a discrete set of velocities and $c_s$ is the sound velocity. The density calculated as $\rho(\vec{x},t) = \sum_i f_i(\vec{x},t)$ followed with the velocity calculation $\vec{u}(\vec{x},t)=\sum_i \vec{c_i}f_i(\vec{x},t)/\rho(\vec{x},t)$.

The hard disks with the radius $R\Delta x$ are immersed in the moving fluid. The mutual interaction force $\vec{g}$ between  fluid and hard disk calculated using the distribution function $f_i (\vec{x}, t)$. The disk boundary is represented as a set of Lagrangian points~\cite{Inamuro-2012} $\vec{X_k}(t + \Delta t)$ and $\vec{U_k}(t + \Delta t)$, $(k = 1, \ldots, N)$ are velocities at those points. The velocities $\vec{u}'$ at the Lagrangian points $\vec{X_k}$ calculated using interpolation

\begin{equation}
	\vec{u}'(\vec{X_k},t+\Delta t) = \sum_{\vec{x}}\vec{u}'(\vec{x}, t+\Delta t)W(\vec{x}-\vec{X_k})
\end{equation}
with summation over all lattice nodes $\vec{x}$, and the weighting function $W$ is defined~\cite{Peskin-1972,Peskin-2002} by	
\begin{equation}
	W(x, y, z)=w\left(\frac{x}{\Delta x}\right)w\left(\frac{y}{\Delta x}\right),
\end{equation}

\begin{equation}
	w(r) = 
	\begin{cases}
		1/8\left(3-2|r|+\sqrt{1+4|r|-4r^2}\right), & |r|\leq 1, \\
		1/8\left(5-2|r|-\sqrt{-7+12|r|-4r^2}\right), & 1\leq |r|\leq 2, \\
		0, & \textrm{otherwise}.
	\end{cases}
\end{equation}

The force $\vec{g}(\vec{x}, t + \Delta t)$ acting on the particle is calculated iteratively in four steps~\cite{Inamuro-2012}

\begin{enumerate}[label=I\arabic*.]
	\addtocounter{enumi}{-1}
	
	\item Initialize value of the force $\vec{g}$ at the Lagrangian points
	\begin{equation}\label{eq:g0}
		\vec{g_0}(\vec{X_k}, t+\Delta t)=\frac{\vec{U_k}-\vec{u'}(\vec{X_k}, t+\Delta t)}{\Delta x},
	\end{equation}

	\item Compute $n$-th iteration $(n = 0, 1,\ldots, l)$ of the force	at the grid points
	\begin{equation}
		\vec{g_n}(\vec{x}, t+\Delta t)=\sum_{k=1}^{N}\vec{g_n}(\vec{X_k}, t+\Delta t)W(\vec{x}-\vec{X_k}),
	\end{equation}
	and correct the velocity at the grid points accordingly	(note, it is not the iteration)
	\begin{equation}
		\vec{u_n}(\vec{x}, t+\Delta t)=\vec{u'}(\vec{x}, t+\Delta t)+\vec{g_n}(\vec{x}, t+\Delta t)\Delta x. 
	\end{equation}

	\item Interpolate the velocity at the Lagrangian points
	\begin{equation}
		\vec{u_n}(\vec{X_k}, t+\Delta t) = \sum_x \vec{u_n}(\vec{x}, t+\Delta t)W(\vec{x}-\vec{X_k}).
	\end{equation} 

	\item Calculate $n-$th iteration of the force
	
	\begin{equation}\label{eq:g5}
		\vec{g}_{n+1}(\vec{X_k}, t+\Delta t)=\vec{g}_n(\vec{X_k}, t+\Delta t)+\frac{\vec{U_k}-\vec{u_n}(\vec{X_k}, t+\Delta t)}{\Delta x}.
	\end{equation}
	
\end{enumerate}

We use the non-dimensional variables following Inamuro's review paper~\cite{Inamuro-2012}. We found the recommendation of reference~\cite{Suzuki-2013} to keep no-slip conditions at the disk surface is working well in our case and use four or five iterations to calculate force, $l = 4-5$.

The hard disk motion simulated with the classical mechanics equations, with the linear velocity $\vec{U}$ evolution
\begin{equation}
	M \frac{d\vec{U}(t)}{dt}=\vec{F}(t)\label{eq:fma},
\end{equation}
and angular velocity evolution
\begin{equation}\label{eq:tiw2}
	\frac{1}{2}MR^2 \frac{d\omega}{dt} = T_z.	
\end{equation}
where $ M$ is the disk mass, $\vec{F}(t) = - \sum\vec{g}(t)$ is the total force acting on the disk, and $T_z$ is the component of the force moment, $\vec{T} = -\sum \vec{r} \times \vec{g}$.

\section{Simulation details and analysis}
\label{sec:simulation}

This section presents the details of the simulations of the hard disk chain movement in the Poiseuille flow. We vary the confinement ratio $\gamma = 2R/W$ of the disk with radius $R$ to the channel width, the number $N$ of hard disks (particles) in the chain, and apply many initial displacements of the particle in the chain — the symmetric displacement of particles along the channel symmetry axis, the random displacement of particles, and displacement of one particle in the transversal to the fluid flow direction. 

\subsection{Disk-chain movement simulation}
Simulations were done mainly using the open-source software Palabos~\cite{Latt-2021} with some modifications in order to incorporate the computational model presented in the previous section. Palabos is a library written in C++ that uses MPI for parallelization. Palabos partitions lattice into smaller regions distributed over the nodes or cores of a parallel machine. Other data types which require a small amount of storage space are duplicated on every node. Chain of particles processing is done as a data processor, which is parallelized. The current problem has been computed with 22 threads on a node with an Intel Xeon Gold 6152 2.1GHz processor with DDR4 2.666 GHz 768GB memory~\cite{HSE-HPC}.

Computational domain is $80\Delta x \times 20000\Delta x$. We perform $1.5 \cdot 10^6$ LBM time steps until the fully developed flow. Boundary conditions on the top and bottom of the channel are walls (bounce-back), and on the left and on the right, the Poiseuille profile is set with an average velocity $u_{av} = 0.015=u_{max}/2$. The channel width is varyed between $W=20\Delta x$ and $W=100\Delta x$, and the channel length $L$ is usually 250 times longer.

After the flow is fully developed, the particles (hard disks) are immersed on the central axis of the channel with the distance between the centers of the particles $a = 4R$. Radius of disks are all equal, $R = 5\Delta x$. The disks in the LBM-IB method are represented by 100 points uniformly distributed around a circle. The number of IB iterations for the force calculation was four and five to check the convergence of iterations.

The particle in the middle of the chain is shifted above the axis to excite oscillations. Fig.~\ref{fig:initial} illustrates the initial position of 8 particles with the particle fifth shifted.  The density of the fluid is $\rho_{fluid} = 1$.   The density of the disks is greater than the density of the surrounding fluid $\rho_{hd} = 1000/840 \approx 1.19$.   We chose this ratio to mimic the ratio of the density of water to the density of oil in the experiment~\cite{BTBZ-2006}. The model uses the two-dimensional D2Q9~\cite{Krueger-book} velocity representation.

We check simulation with the code written in python. The reason was to correctly implement the boundary conditions for the fluid in the channel and correctly incorporate the bounce-back boundary conditions along the channel with the boundary conditions at the incoming and outcoming fluid flow corresponding to the Poiseuille velocity profile. Simulations stop while the rightmost particle reaches the outcoming boundary.

\subsection{Chain oscillation analysis}

For the chain oscillation spectrum, we adopt the same method as in the analysis of the experimental data in Ref.~\cite{BTBZ-2006} using the coordinate system shown in Fig.~\ref{fig:initial}. Oscillations in the direction transversal to the flow are calculated using deviation of the particle $n$ $(i = 1, 2, \ldots, N )$ from the channel axis, i.e. it is oscillations around the axis, $y_n(t_i)$ at the discrete moment of simulation $t_i$ $(i = 1, 2, \ldots, T )$. Longitudinal oscillations are calculated from the difference $\xi_n(t)$ of the neighboring particles

\begin{equation}
	\xi_n(t_i ) = 
	\begin{cases}
		0,& n=0 \\
		x_n( t_i )-x_{n-1}(t_i ), & n \in [1, N-1].
	\end{cases}
	\label{eq-ksi}
\end{equation}
and distances $\xi$ is connected to the relative oscillations of the neighboring particles~\cite{BBZT-2007}.

The discrete Fourier transform performed with NumPy library~\cite{Harris-2020} using expressions~\cite{BTBZ-2006} (we drop $\Delta x$ and $\Delta t$ in order to shorten expressions)

\begin{equation}
	\tilde{X}(k,\omega)=\sum_{n=1}^{N}\sum_{t=1}^{T}\xi(n, t)e^{-(2\pi i/N)(k-1)(n-1)}e^{-(2\pi i /T)(\omega-1)(t-1)} 
	\label{eq-Xk}
\end{equation}

and
\begin{equation}
	\tilde{Y}(k,\omega)=\sum_{n=1}^{N}\sum_{t=1}^{T}y(n, t)e^{-(2\pi i/N)(k-1)(n-1)}e^{-(2\pi i /T)(\omega-1)(t-1)}
	\label{eq-Yk}
\end{equation}

\section{Single particle in the Poiseuille flow}
\label{sec-single}

In the following, we will use the dimensionless variables, and all LBM variables are normalized by the particle radius $R\Delta x$ and by the maximum velocity $u_{max}$ of Poiseuille flow. The normalized variables are denoted with a tilde. Therefore, the particle radius $\tilde{R}{=}1$ and maximum velocity  $\tilde{u}_{max}{=}1$.

\begin{figure}
	\begin{center}
		\includegraphics[width=.9\columnwidth]{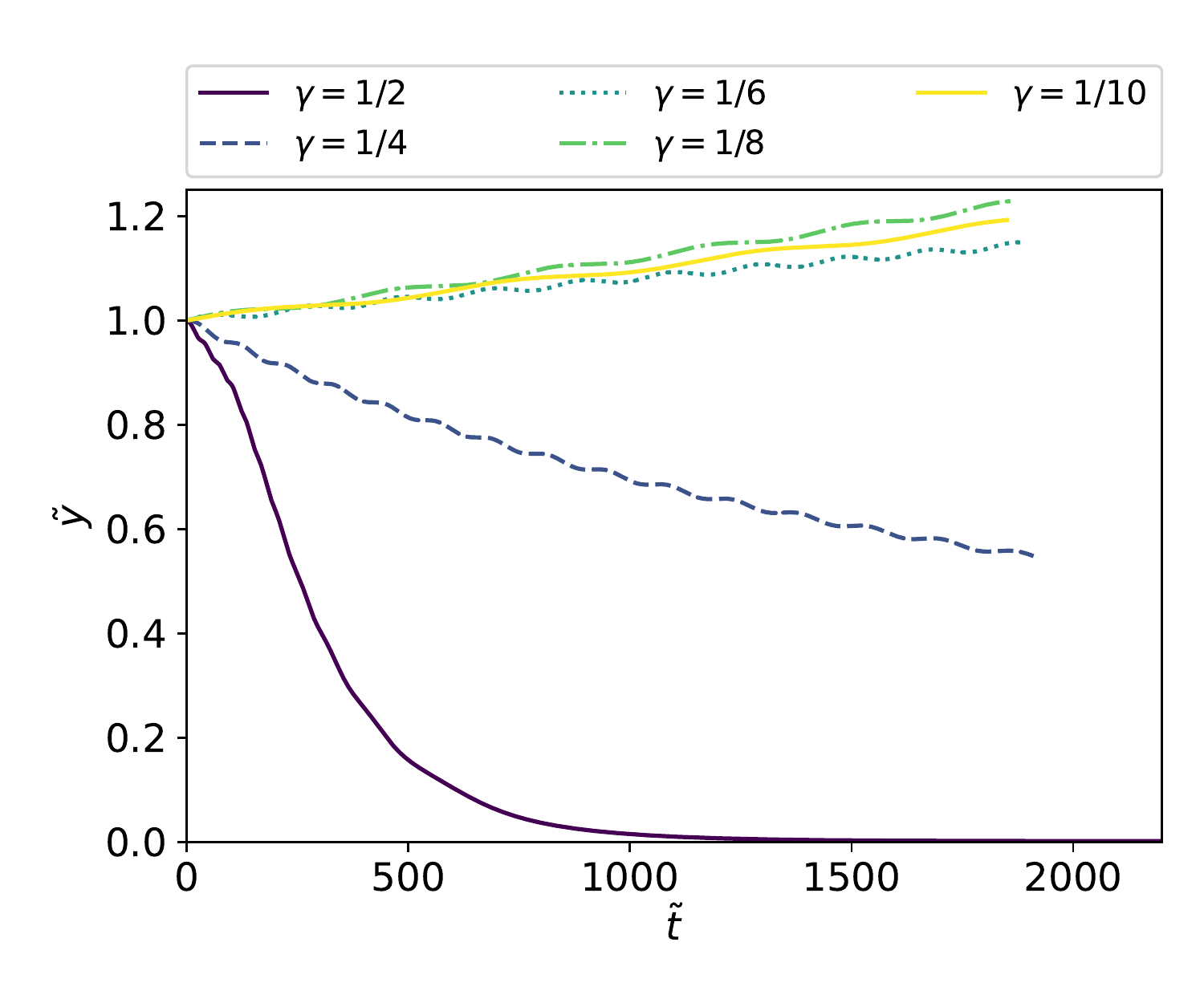}
		\caption{Position $\tilde y$ with time $\tilde t$ of the particle transversal migration in the channel with different confinement ratios $\gamma = 1/2, 1/4, 1/6, 1/8$, and $1/10$.}
		\label{fig:one}
	\end{center}
\end{figure}

\begin{figure}
	\begin{center}
		\includegraphics[width=.9\columnwidth]{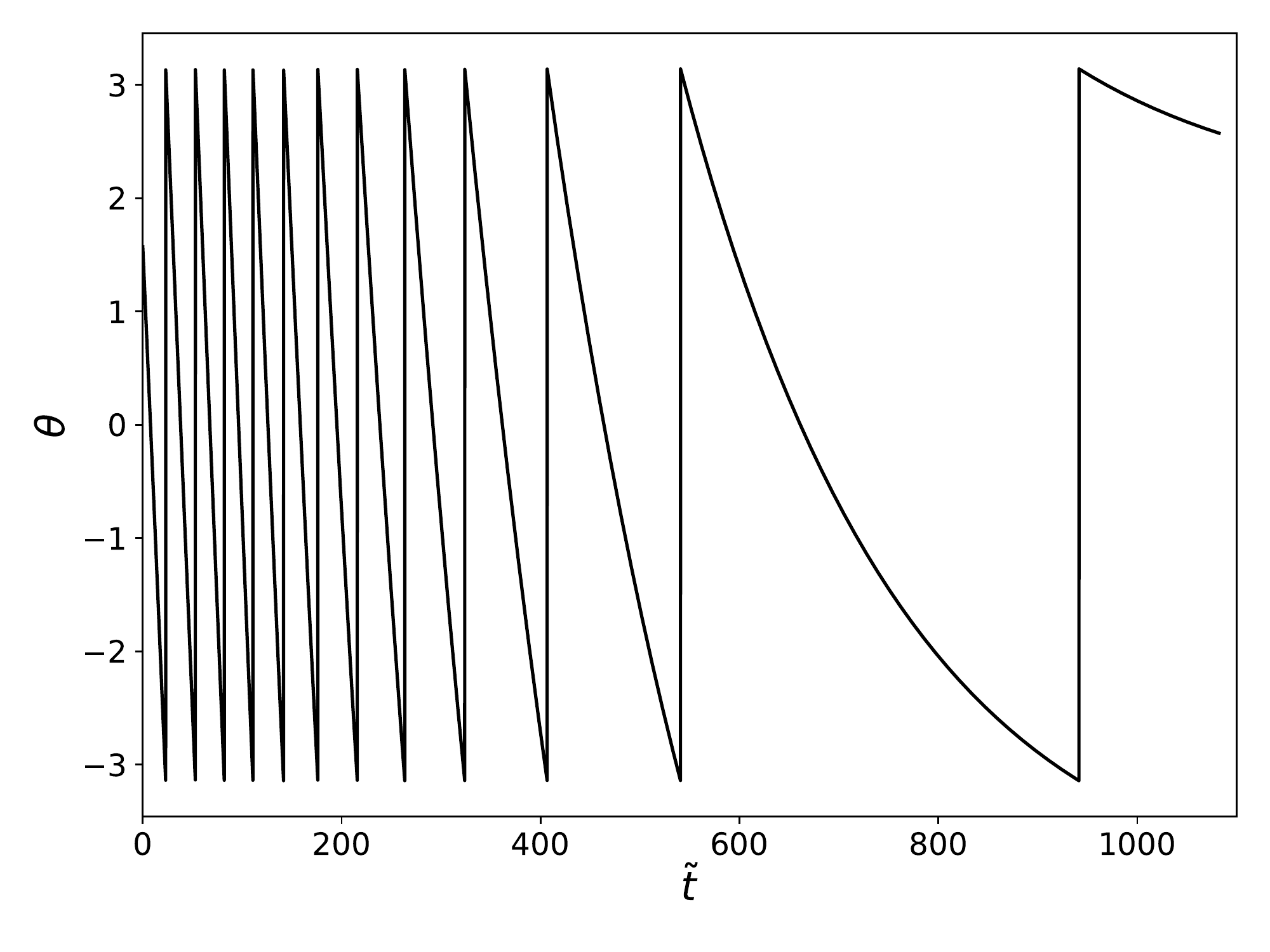}
		\caption{Rotation of the hard disk -- angle dependence of the boundary point in time. Confinement ratio $\gamma=1/2$.  }
		\label{fig:one_rotation}
	\end{center}
\end{figure}

As we already discussed, the particle experiences lifting force due to the velocity gradient of Poiseuille flow. 
Indeed, it is the case of our simulations as shown in Fig.\ref{fig:one}. The particle of radius $\tilde R=1$ is placed apart from the axis at a distance $\tilde{y}_0 = 1$ in the channels with width $\tilde{W} = 4, 8, 12, 16$, and $20$.
During the simulation, the particle migrates to the channel axis for large confinement ratio $\gamma = 2\tilde R/\tilde W=1/2$ and $1/4$, and to the equilibrium position away from the channel axis for smaller values of confinement ratio (not all cases converge to the equilibrium position reached within the time of the Fig.~\ref{fig:one}). We must note that the time scale of migration is larger than $1000$. The particle is rotating in the migration process, and in Fig.~\ref{fig:one_rotation} we present the change in time of the rotation angle for the narrow channel $\gamma=1/2$. The typical time scale of the hard disk rotation is an order of magnitude shorter than the time of migration, and the rotation period $\tilde T$ is presented in Table~\ref{tab1}.

\begin{figure}
	\begin{center}
		\includegraphics[width=.9\columnwidth]{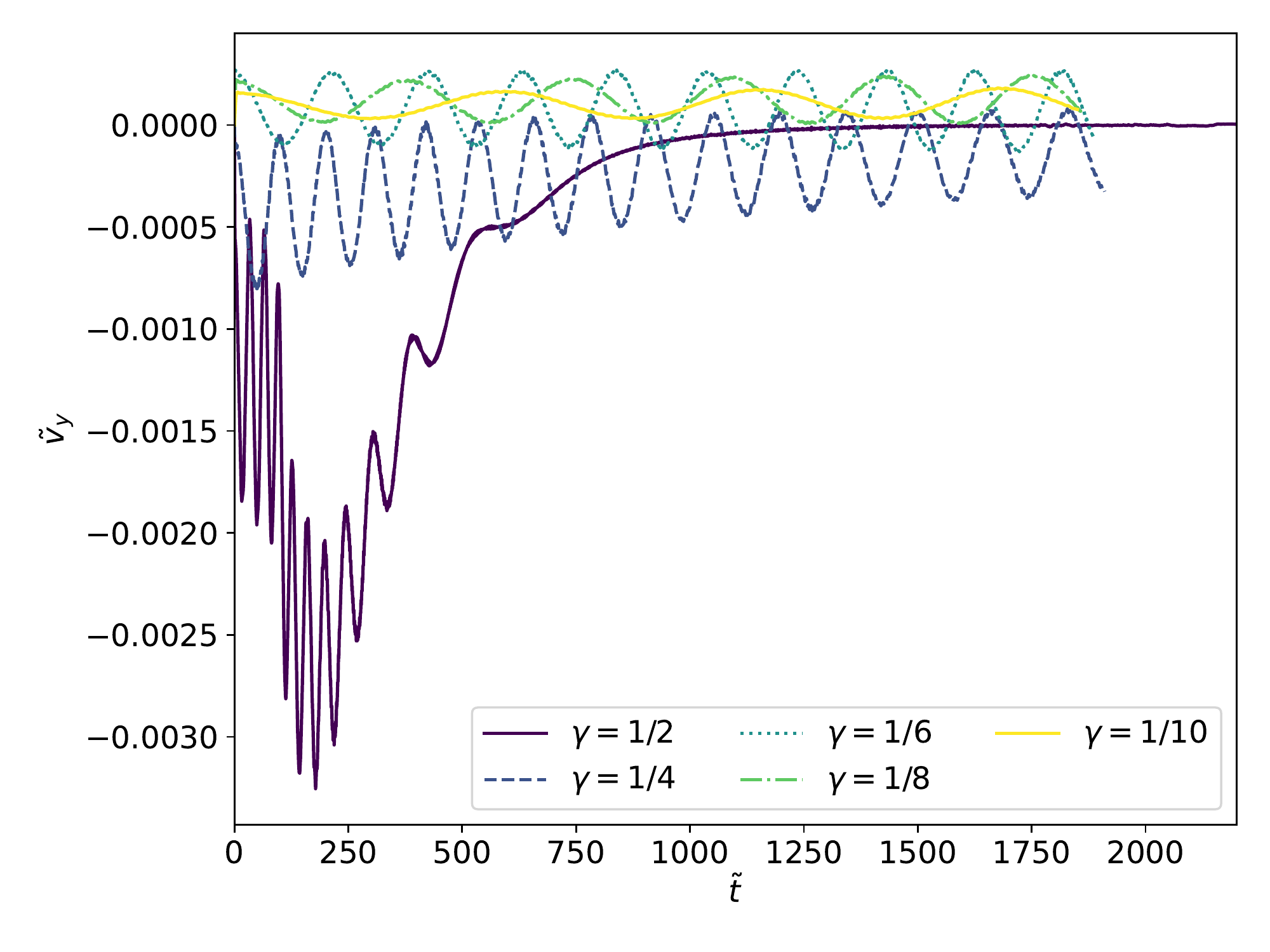}
		\caption{Variation of the particle transversal component of velocity $\tilde{u}_y$ in time for different values of confinement ratio $\gamma$. }
		\label{fig:one_velocities}
	\end{center}
\end{figure}

\begin{table}
\begin{tabular}{|r|r|r|r|}
	\hline
	$\gamma$ & $\tilde T$ & $\tilde \Omega$ & $\tilde u_{rel}$ \\
	\hline
	
	1/2 & 54.80 & 0.1147 & -0.1287 \\ 
	\hline
	
	1/4 & 133.16 & 0.0472 & -0.0333 \\ 
	\hline
	
	1/6 & 200.19 & 0.0314 & -0.0128 \\ 
	\hline
	
	1/8 & 343.57 & 0.0183 & -0.0101 \\ 
	\hline
	
	1/10 & 546.27 & 0.0115 & -0.0054 \\ 
	\hline
\end{tabular}
\caption{\label{tab1}The period $\tilde T$ of the particle oscillations, angular velocity $\tilde \Omega$, and relative velocity of the particle $\tilde u_{rel}$.}
\end{table}

\begin{figure}
	\begin{center}
		\includegraphics[width=.9\columnwidth]{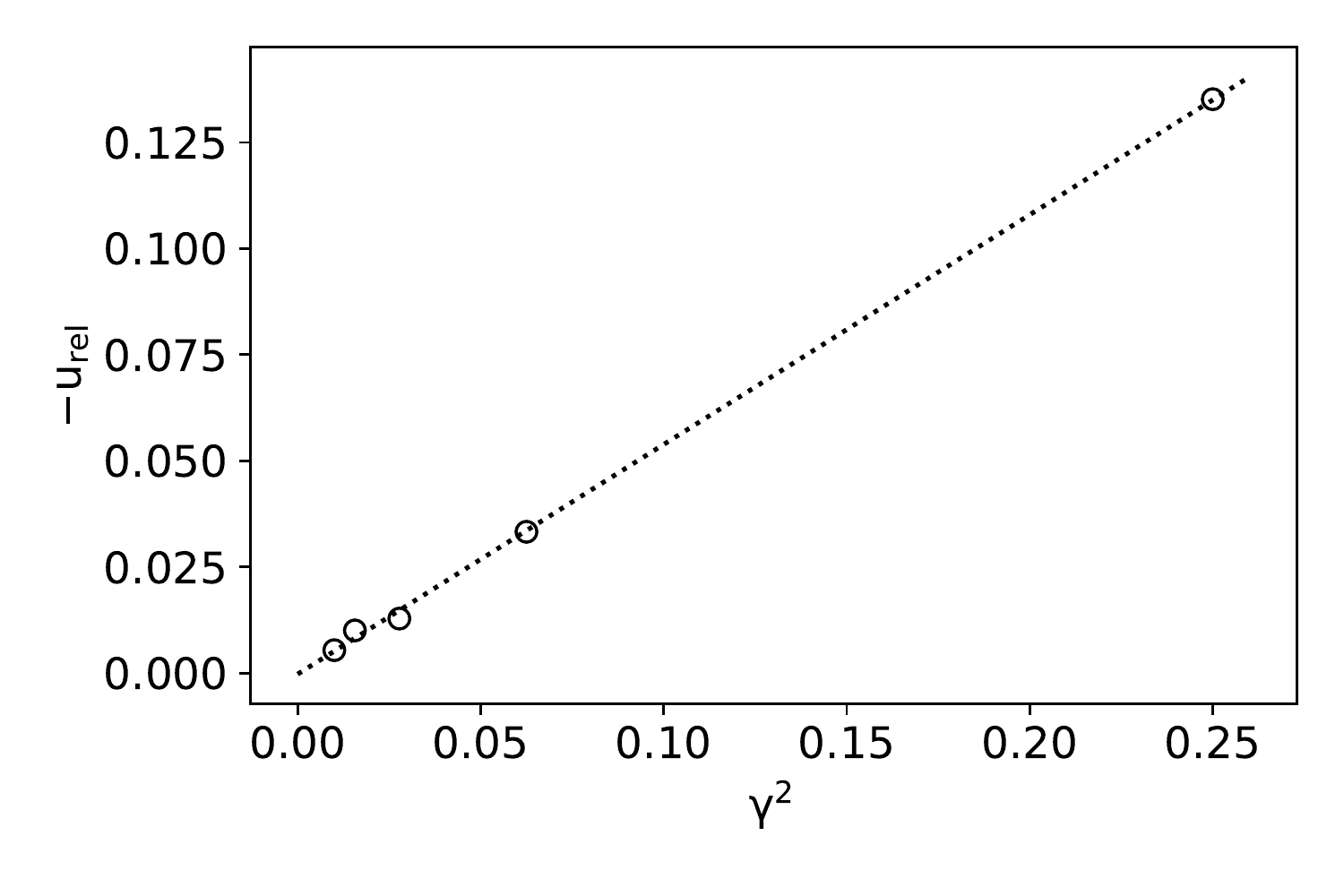}
		\caption{Relative particle velocity $u_{rel}$ as function of the square of confinement ratio  $\gamma$. The dotted line is the fit to the data.}
		\label{fig:relative}
	\end{center}
\end{figure}

Table~\ref {tab1} entries present the period $\tilde T$ of single-particle oscillations in the channels, the corresponding angular velocity $\tilde \Omega$, and the relative velocity of the particle $\tilde u_{rel}$ depending on the confinement ratio $\gamma$ value. We have to note that the narrow channel the larger particle retardation, i.e., the velocity growth with the larger confinement ratio $\gamma$ reflecting the wall proximity. 
The relative velocity of the particle $\tilde u_{rel}$ is shown in the figure~\ref{fig:relative}, and plotted as a function of the square of the confinement ratio. Fit to the data gives $\tilde u_{rel}\approx 0.51 \gamma^2$, with $\gamma$ dependence similar to the case calculated by Simha~\cite{Simha-1936} for the immersed infinite cylinder. We have to note that the numerical value of the coefficient behind $\gamma^2$ obtained analytically does depend on the assumptions used for the treatment of the equation solving (f.e., see discussion section in Ref~\cite{Saffman-1965}). 

We have to note that the relative velocity $\tilde u_{rel}$ and angular velocity $\tilde \Omega$ are about the same for the aspect ratio $\gamma=1/2$ and relative velocity decays faster with the decreasing confinement ratio $\gamma$ than the angular velocity. The combination of rotation and transversal migration leads to the oscillation of the transversal velocity as shown in Fig.~\ref{fig:one_velocities}, from which one can estimate the frequency of oscillations depending on the channel confinement ratio $\gamma$. According to analytical estimations~\cite{Simha-1936, Rubinow-1961},  the frequency is associated with the proximity to the wall and the shear velocity gradient.

More details on the simulation of the lonely particle in two-dimensional Poiseuille flow with small Reynolds numbers will published elsewhere~\cite{Future-1}. The steady-state positions of the sphere in the tube was simulated recently for the higher Reynolds numbers than we consider in the paper~\cite{Fox-2021}.

\section{Emergence of collective spectrum}
\label{sec:spectrum}

This section presents the results of our simulations using the LBM-IB approach described above. We found some ``critical'' number of particles in the chain at which the collective spectrum of the moving particles spectrum appears. 

The spectrum of ``flowing crystal''~\cite{BBZT-2012} was first reported in the pioneering work~\cite{BTBZ-2006}. In the experiment~\cite{BTBZ-2006}, the water droplets were injected into the oil moving between two plates. The plates squeezed the droplets, so the geometry was quasi-two-dimensional, and the velocity of droplets was slower than the velocity of the surrounding oil. The velocity difference leads to the perturbation of the oil, which mediates a drag force on the other droplets. For the quasi-two-dimensional geometry of the experiment, the hydrodynamic potential of the single drop at the large distances $\vec r$ will approximately satisfy the two-dimensional Laplace equation with solution $\phi \approx R^2 u_{rel}\vec{r}/r^2$ which is the potential of the two-dimensional dipole~\cite{Cui-2004}, with $u_{rel}$ is the drop velocity relative to the oil flow. The spectrum of the dipoles chain was calculated analytically and compared well with the spectrum of about 60 water droplets injected into the oil.

The Poiseuille flow is axisymmetric and non-homogeneous in the direction transversal to the axis. It is known from early  experiments~\cite{Segre-1961,Goldsmith-1961,Karnis-1963} that particles can migrate axially in the Poiseuille flow. 
The theoretical considerations suggest that sphere in 3D shear flow and disk in 2D shear flow may seem to rotate~\cite{Tollert-1954} and may experience lifting force~\cite{Segre-1961,Rubinow-1961,Saffman-1965} pointing perpendicular to the streamlines. The lifting force is a function of the shear gradient, particle radius, viscosity, density, and relative velocity~\cite{Rubinow-1961,Saffman-1965}. There are analytical calculations~\cite{Simha-1936} that suggest that the relative velocity of the particle is proportional to the square of the confinement ratio multiplied by the value of the Poiseuille flow at the axis, $u_{rel}\propto \gamma^2u_{max}$. This relative velocity will produce the dipole-like potential $\phi \propto R^2 \gamma^2u_{max}\vec{r}/r^2$

So, one can expect individual particles in the chain to move axially and transversally and rotate due to the Poiseuille flow and proximity to the wall. The perturbed by particles fluid will mediate the interaction between particles in the chain, and the collective excitations will emerge. The spectrum should be similar to those obtained in the experiment with the drops squeezed between plates~\cite{BTBZ-2006}.
In addition, one can expect the spectrum intensity would be higher with the larger confinement ratio due to the larger relative velocity of particles and, therefore, the stronger particle interaction.

In the following, we will use the dimensionless variables, and all LBM variables are normalized by the particle radius $R$ and by the maximum velocity $u_{max}$ of Poiseuille flow. The normalized variables denoted with tilde; therefore, the particle radius $\tilde{R}{=}1$ and maximum velocity $\tilde{u}_{max}{=}1$.

\begin{figure}
	\begin{center}
		\includegraphics[width=1\columnwidth]{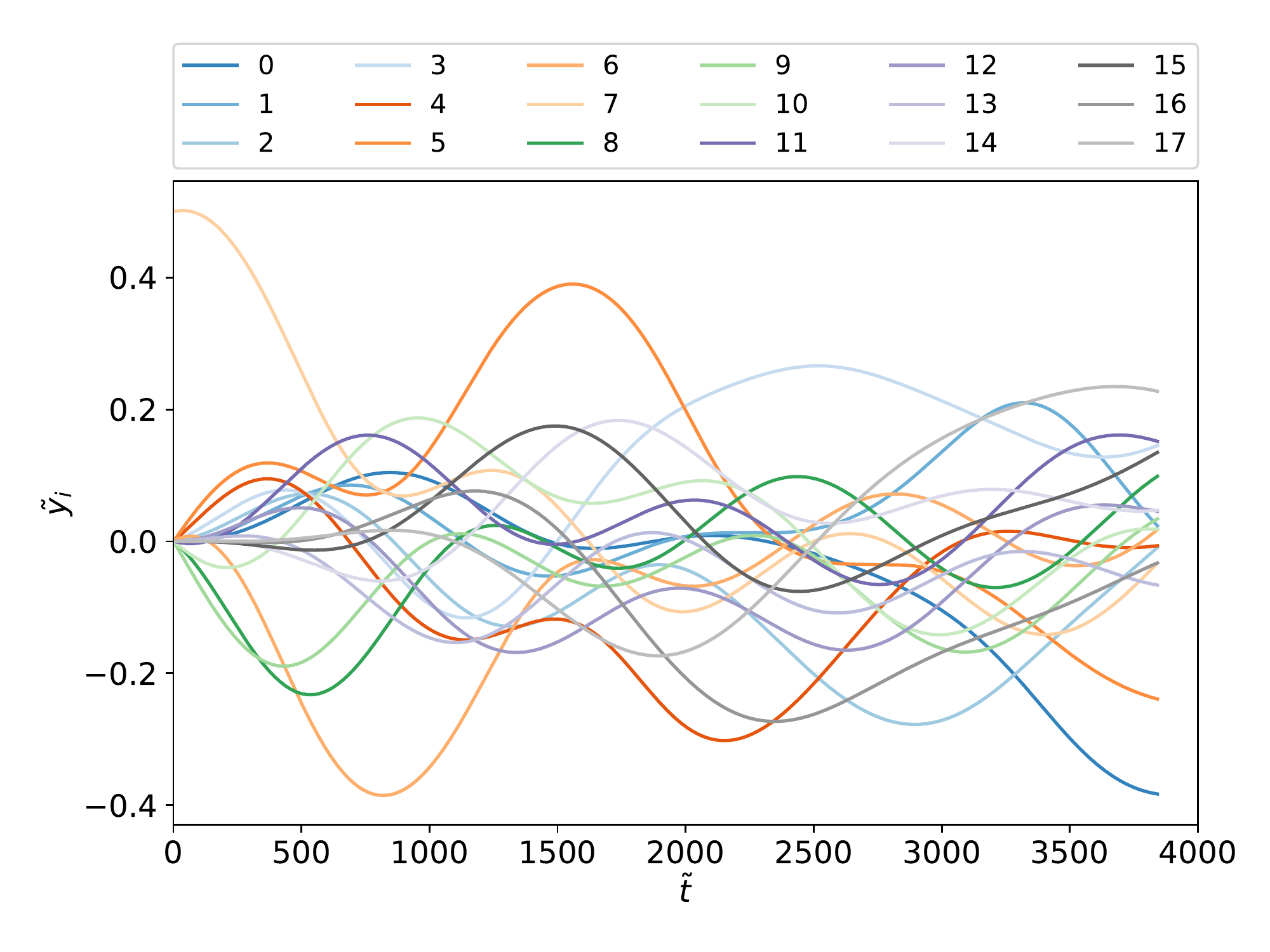}
		\caption{Transversal oscillations of $N=18$ particles in the chain - the transversal coordinate $\tilde y_i$ of the $i$-th particle, $i=0,1,\ldots,N-1$. }
		\label{fig-y-chain}
	\end{center}
\end{figure}

\begin{figure}
	\begin{center}
		\includegraphics[width=1\columnwidth]{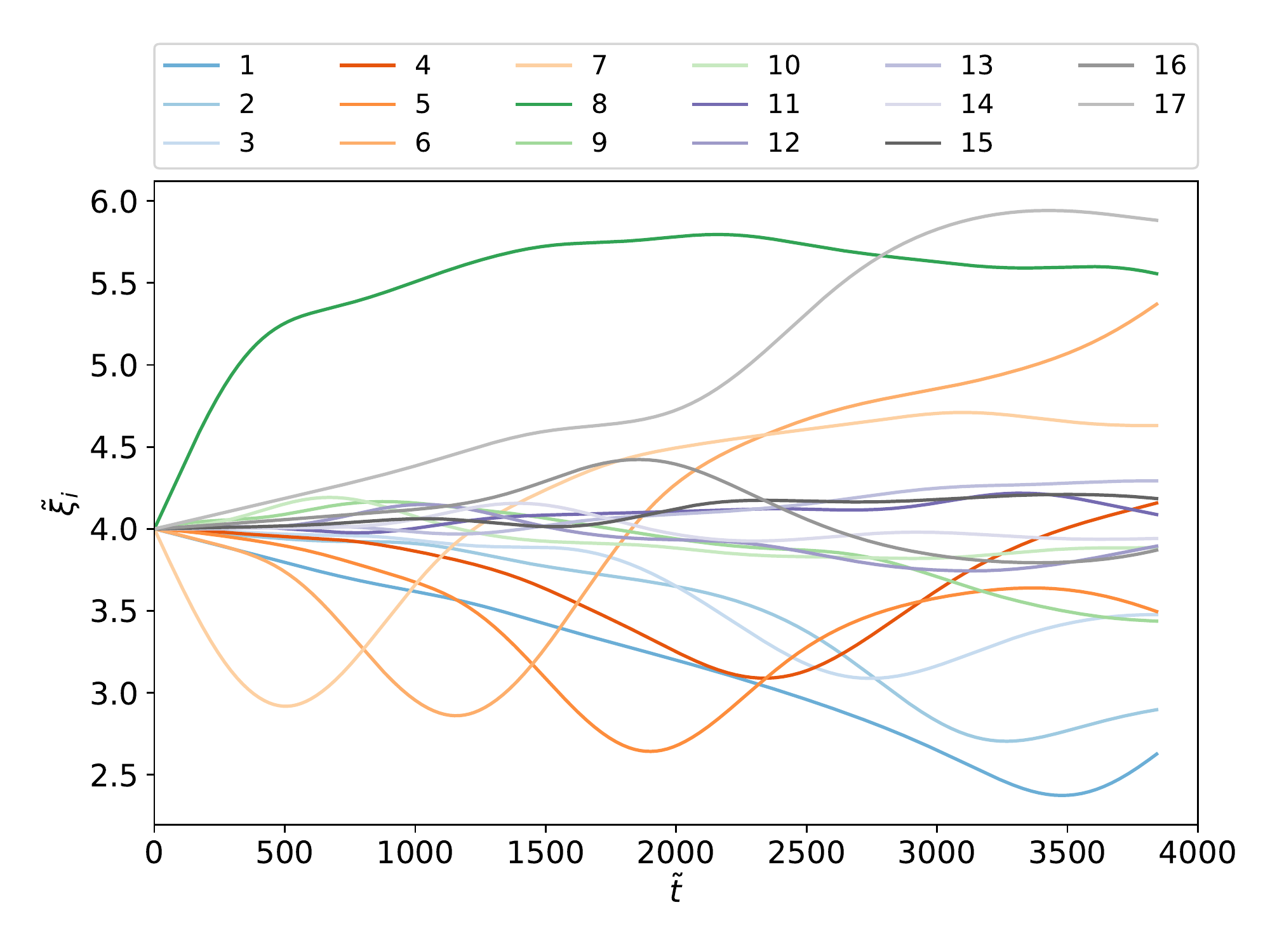}
		\caption{Relative oscillations of $N=18$ particles in the chain - the difference $\tilde\xi=\tilde x_{i}-\tilde x_{i-1}$ of the longitudinal positions $x_i$ of the particles, $i=1,\ldots,N-1$.}		
		\label{fig-ksi-chain}
	\end{center}
\end{figure}

The geometry of the computer experiment is shown in Fig.~\ref{fig:initial}. We simulate chain with $N$ particles of dimensionless radius $R=1$ with the distance between centers $a=4$ (four times wider than radius $R$) in the channel with confinement ratio $\gamma$ under the Poiseuille flow with the dimensionless maximum velocity $\tilde u_{max}=1$. It is the case of viscous flow with the low Reynolds number $Re=0.25$. 

The $N$ particles placed along the channel axis with the middle one shifted transversally with distance $\tilde y_0$, thus breaking the symmetry of the chain and triggering particle oscillations. We also used the small random displacement of the particles from the perfect positions in the one-dimensional lattice and found qualitatively similar results and noisy ones. We simulate chain movement in dimensionless time up to several hundred and collect the position of all disk centers in the transversal direction and the difference between neighboring particle centers in projection to the x-axis. The collected data we use for the spectrum calculation using Expr.~(\ref{eq-Xk}-\ref{eq-Yk}). The transversal positions $\tilde y_i$ of the $N=18$ particles is shown in Fig~\ref{fig-y-chain} and the relative distances of particles $\tilde \xi$  (Expr.~\ref{eq-ksi}) is shown in Fig~\ref{fig-ksi-chain}. One can note that the frequencies of the oscillations are comparable with the frequencies generated by the lifting force and rotating force discussed in the previous subsection for the case of one-particle movement. At the same time, we must note that the initial disturbance of just one particle in the chain produces oscillations of all $N$ particles in the chain. The new phenomena appear for a long enough chain -- the collective spectrum.

\begin{figure}
	\begin{center}
		\includegraphics[width=1\columnwidth]{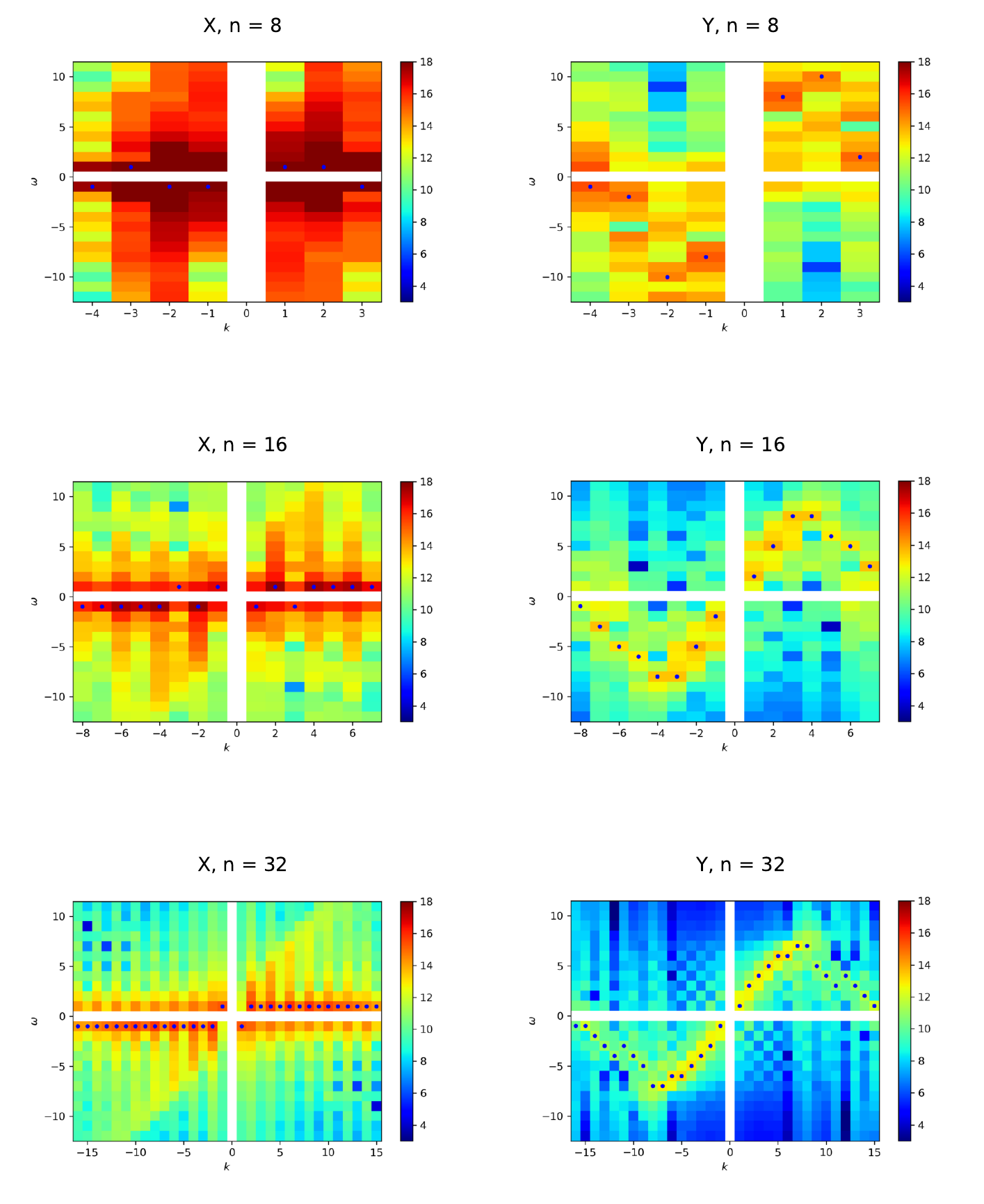}
		\caption{Intensity plot of the power spectrum in the narrow channel with confinement ratio $\gamma=1/2$ and a different number of particle $N$ in the chain. Left column – power spectrum of the longitudinal oscillations. Right column - the power spectrum of the transversal oscillations. From top to bottom $N=8,16$ and 32. }
		\label{N-spectrum}
	\end{center}
\end{figure}

Figure~\ref{N-spectrum} presents the emergence of the collective spectrum in the disk chain with the increasing number of particles $N$ in the chain. It is visible from the figure that the collective spectrum appears with the increasing number of particles in the chain. The intensity plot of the power spectrum shows the emergence of the visible dispersion dependence for the number of the particle about $N\approx10$ (not shown in the figure) and become well pronounced and sharper already with $N=16$. 

The longitudinal dispersion law is quite different from those in the quasi-two-dimensional case, which is associated with the dipole-like potential flow of fluid around the retarded drop - it is the linear spectrum in the left column of Figure~\ref{N-spectrum}.

The difference of the obtained disk chain spectrum in the two-dimensional narrow channel with the spectrum of squeezed drop chain in the quasi-two-dimensional geometry is as follows. The main contribution to the drop interaction is through the dipole-like potential for squeezed case, in which the relative drop velocity is quite large and the Reynolds number in the experiment ($Re<10^{-2}$) is relatively low. In our case, the disks in the narrow channel with a more significant but still small Reynolds number $Re=1/4$ are due to the proximity to the wall and the velocity gradient of the Poiseuille flow. The relative velocity of the particles is much smaller than the average fluid velocity, and dipole-like potential does not affect the spectrum much. We simulate chain for the number of confinement ratio to support that point, and indeed, the spectrum becomes sharper for the narrow channel and larger confinement ratio.

The open question remains -- it is the existence of the gap in the transversal dispersion law, the dependence of frequency $\omega_y$ from the wavenumber at the small values. The absence of the translational invariance of the chain in the transverse direction $y$ can lead to a gap in the spectrum. 
The possible value of the gap can be estimated with the following considerations. For example, if one shifts the infinite chain from the axis at some distance $y_0$ in the narrow channel $\gamma=1/2$, the chain will relax to the steady position at the channel axis. It is about the same as the relaxation time of the single-particle shown by the lower line in the ~\ref{fig:one}. We check in simulations for the single-particle using several values of initial displacement $y_0\in [0.25:1]$, the distance $\tilde y(t)$ decay exponentially $\tilde y \approx \tilde y_0\exp(-\tilde t/\tilde T_{relax})$ with the $\tilde T_{relax}\approx 200$. This value of the relaxation period is shorter than the visible period of the oscillations of the individual particles shown in the fig.~\ref{fig-y-chain}. Therefore, the effective frequency of relaxation $\tilde \omega_{relax} \propto 1/\tilde T_{relax}$ is larger than some frequencies of the chain oscillations shown in the figure. This fact contradicts our assumption that the gap in the spectrum relates to the time of the chain relaxation to the steady position. Simulations of the very long chain detecting accurate estimation of the dispersion law are necessary to associate the transversal spectrum with the optical spectrum or the gapless acoustic phonon spectrum.

\section{Discussion}
\label{sec:discussion}
The spectrum of collective excitations of flowing particles was found in the pioneering research by Weizmann group~\cite{BTBZ-2006}. They use the quasi-two-dimensional oil flow with water droplets squeezed by two plates. Water droplets of equal radii were injected in the moving oil at a constant rate with equal inter-droplet distances. The droplet velocity was smaller than the surrounding oil due to the friction between droplets and plates. The motion of droplets was tracked using the microscope and digital camera, and the spectrum was calculated with a pack of about 60 droplets. The spectrum reflects the collective oscillations of the drops in the chain, and authors argued the drop interaction is due to the dipole-like potential originating from the difference of the drop velocity relative to the fluid, which in turn is due to the friction of the squeezed drops at the top and bottom plates. The analytical arguments well support this picture~\cite{BBZT-2012} as well as following numerical experiments from other groups.  For example, Ref.~\cite{Liu-2012} authors simulate drop interaction using the dipole potential explicitly and reproduce well the spectrum of Ref.~\cite{BTBZ-2006}.

In this paper, we report the auxiliary possibility for the collective spectrum of the moving chain. We simulate buoyancy hard disks neutrally in the two-dimensional channel. The disks were immersed in the Poiseuille flow of the surrounding fluid. We argue in the paper that under the flow with Reynolds number slightly less than unity, the main contribution to the particle interaction is due to the gradient of Poiseuille velocity and the proximity of particles to the wall. The relative velocity of particles does produce the dipole potential, but the velocity value is too small to produce detectible dipole interaction of the particles. The collective spectrum is just the plane waves in the longitudinal direction and a combination of the direct and reflected from the walls plane waves in the transversal direction.

In conclusion, we numerically found the collective spectrum of the neutral buoyancy particle chain in the narrow two-dimensional channel at the moderate values of Reynolds number.

\begin{acknowledgments}
The authors are thankful to E. Burovski and V. Shchur for discussing the numerical results and careful reading of the manuscript. 
LS acknowledges the helpful discussion of the power spectrum with I. Kolokolov and V. Lebedev.
Research supported by the grant 19-11-00286 of the Russian Science Foundation. 
The simulations were done using the computational resources of HPC facilities at HSE University.      
           
\end{acknowledgments}

\end{document}